\documentclass[showpacs,aps,twocolumn]{revtex4}
\usepackage{epsfig}
\usepackage{graphicx}
\usepackage{amsmath,amssymb,amsfonts}
\usepackage{array}
\usepackage{url}
\usepackage{multirow}
\usepackage{float}
\usepackage{lineno}
\usepackage{xspace}
\usepackage{ulem}
\usepackage{amsmath}
\usepackage{hyperref}
\usepackage{tikz}
\usepackage{graphicx}

\newcommand{\bef}{\begin{figure}}
\newcommand{\eef}{\end{figure}}
\newcommand{\bc}{\begin{center}}
\newcommand{\ec}{\end{center}}

\newcommand{\be}{\begin{equation}}
\newcommand{\ee}{\end{equation}}
\newcommand{\bea}{\begin{eqnarray}}
\newcommand{\eea}{\end{eqnarray}}

\def\ba{\begin{eqnarray}}
\def\ea{\end{eqnarray}}

\begin{document}
\title{Understanding the medium-like effects in the jet-like yield in pp and p--Pb collisions using event generators}

\author{Dushmanta Sahu}
\author{Antonio Ortiz}
\author{Victor Campos}
\affiliation{Instituto de Ciencias Nucleares, UNAM, Apartado Postal 70-543, Coyoacán, 04510, México City, México}

\begin{abstract}
To understand the dynamics of jet-medium interaction in small systems such as proton-proton (pp) and proton-lead (p--Pb) collisions at $\sqrt{s_{\rm NN}}$ = 5.02 TeV, particle production is studied in three distinct topological regions defined with respect to the charged particle with the highest transverse momentum in the event ($p_{\rm T}^{\rm trig}$). The jet-like yield is defined by the particle density in the toward region ($|\Delta\varphi|<\pi/3$) after subtracting that in the transverse region ($\pi/3<|\Delta\varphi|<2\pi/3$). The activity on the transverse side is used as a proxy for medium-like effects. Three different Monte Carlo event generators --\textsc{Pythia8}, a multiphase transport (AMPT) model, and EPOS4-- are employed to investigate particle yields as a function of $p_{\rm T}^{\rm trig}$ in the interval 0.5-20\,GeV/$c$. Calculations are performed for the $p_{\rm T}$ threshold of 0.5 GeV/$c$ at mid-rapidity ($|\eta| < 0.8$). The jet-like yield in the toward region for pp collisions show interesting dynamics; they are significantly affected by the medium-like effects in the low to intermediate $p_{\rm T}^{\rm trig}$ ($<8$\,GeV/$c$) which is studied through color reconnection and hydrodynamics in \textsc{Pythia8} and EPOS4, respectively. However, the results from AMPT show that the jet-like yield is medium-like modified throughout the entire $p_{\rm T}^{\rm trig}$ range. The jet-like yield in p--Pb collisions using AMPT is also studied. Notably, a dip structure that is observed in the jet-like signal ratio of pp to p--Pb at low $p_{\rm T}^{\rm trig}$ in ALICE data, is reproduced by AMPT model with string melting on, pointing to possible medium-like behavior in small systems. The results of this article also underscore the importance of high-$p_{\rm T}^{\rm trig}$ ($p_{\rm T}^{\rm trig} >$  8 GeV/$c$) for minimizing underlying event biases in jet-related studies.

\end{abstract}
\maketitle
\section{Introduction}

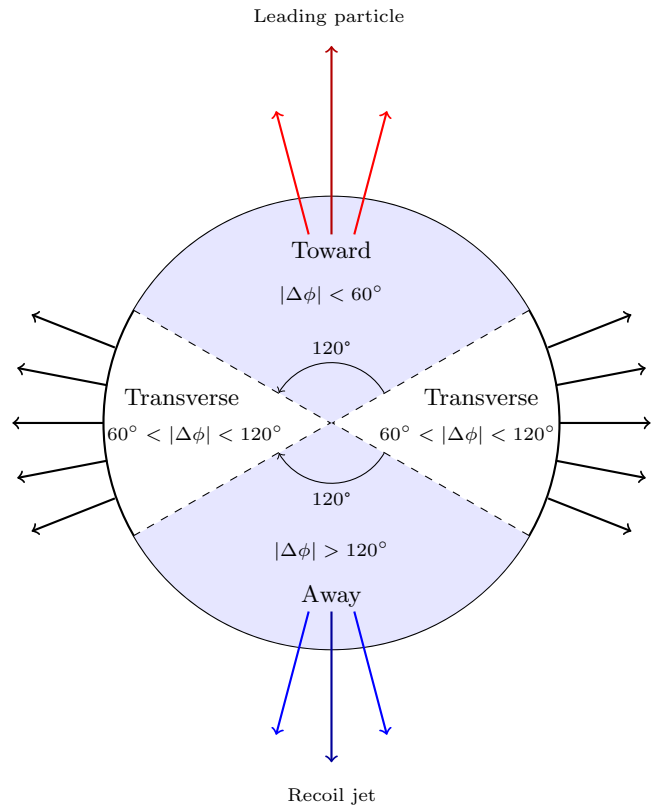
\begin{figure}[h]
\centering
\begin{tikzpicture}[scale=1.0]

\draw[thick] (0,0) circle (3cm);

\draw[thick, dashed] (30:3) -- (-150:3);
\draw[thick, dashed] (-30:3) -- (150:3);

\fill[blue!10] (0,0) -- (30:3) arc (30:150:3) -- cycle;  
\fill[blue!10] (0,0) -- (-30:3) arc (-30:-150:3) -- cycle; 

\draw[->] (30:0.8) arc (30:150:0.8) node[midway, above] {\scriptsize 120°};
\draw[->] (-30:0.8) arc (-30:-150:0.8) node[midway, below] {\scriptsize 120°};

\draw[red, thick, ->] (-0.3,2.5) -- (100:4.2);
\draw[red!70!black, thick, ->] (0,2.5) -- (90:5);
\draw[red, thick, ->] (0.3,2.5) -- (80:4.2);

\draw[blue, thick, ->] (-0.3,-2.5) -- (-100:4.2);
\draw[blue!60!black, thick, ->] (0,-2.5) -- (-90:4.5);
\draw[blue, thick, ->] (0.3,-2.5) -- (-80:4.2);

\draw[black, thick, ->] (-3,0) -- (180:4.2);
\draw[black, thick, ->] (-2.95,-0.5) -- (190:4.2);
\draw[black, thick, ->] (-2.95,0.5) -- (170:4.2);
\draw[black, thick, ->] (-2.85,1) -- (160:4.2);
\draw[black, thick, ->] (-2.85,-1) -- (200:4.2);

\draw[black, thick, ->] (3,0) -- (0:4.2);
\draw[black, thick, ->] (2.95,0.5) -- (10:4.2);
\draw[black, thick, ->] (2.95,-0.5) -- (350:4.2);
\draw[black, thick, ->] (2.85,-1) -- (340:4.2);
\draw[black, thick, ->] (2.85,1) -- (20:4.2);

\node at (10:2) {\small Transverse};
\node at (170:2) {\small Transverse};
\node at (90:2.3) {\small Toward};
\node at (90:1.7) {\scriptsize $|\Delta\phi| < 60^\circ$};
\node at (-90:2.3) {\small Away};
\node at (-90:1.7) {\scriptsize $|\Delta\phi| > 120^\circ$};

\node at (355:1.8) {\scriptsize $60^\circ < |\Delta\phi| < 120^\circ$};
\node at (185:1.8) {\scriptsize $60^\circ < |\Delta\phi| < 120^\circ$};

\node[black, right] at (102:5.5) {\scriptsize Leading particle};
\node[black, right] at (-98:5.0) {\scriptsize Recoil jet};

\end{tikzpicture}
\caption{Illustration of the toward ($|\Delta\phi| < 60^\circ$), transverse ($60^\circ < |\Delta\phi| < 120^\circ$), and away ($|\Delta\phi| > 120^\circ$) regions in the azimuthal plane with respect to the leading particle direction, showing example particles in toward, transverse and away regions.}
\label{fig1}
\end{figure}

In high-energy physics, pp and heavy-ion (Pb--Pb) collisions at facilities such as the Large Hadron Collider (LHC) provide a unique environment to study the fundamental properties of Quantum Chromodynamics (QCD) matter. These collisions produce a wide spectrum of final-state particles, ranging from high-energy jets to low-momentum (soft) hadrons, offering insight into the dynamics of the strong interaction and the possible formation of extreme states of matter such as the strongly interacting quark--gluon plasma (QGP). One of the key challenges in analyzing such collisions lies in disentangling contributions from the primary hard-scattering processes, where high-momentum incoming partons interact to produce jets, from the softer and more diffuse particle production that constitutes the underlying event (UE). In high-energy pp collisions, the UE arises from the complex interplay of several components, such as beam remnants, semi-hard multi-parton interactions (MPI), and initial- and final-state radiation (ISR/FSR)~\cite{Sjostrand:1987su}. These mechanisms act in parallel with the primary hard scattering, contributing significantly to the final-state particle distributions. Thus, characterizing the UE is essential for a complete understanding of hadronic collision dynamics, especially in disentangling the hard processes from background activity. To isolate and study the UE, one usually divides the azimuthal plane relative to a high-$p_{\rm T}$ leading particle into distinct regions~\cite{CDF:2001onq}. As illustrated in Fig.~\ref{fig1}, the event is segmented into three topological regions: (i) the \textit{toward} region, containing the leading jet; (ii) the \textit{away} region, which typically contains the recoil jet; and (iii) the \textit{transverse} region, which is orthogonal to the jet axis and most sensitive to UE activity. By analyzing observables such as charged-particle yields and $p_{\rm T}$ spectra in these regions, UE contributions can be studied independently of the hard-scattering process.

Traditionally, small systems such as pp collisions have been used as a baseline for heavy-ion physics, under the assumption that they lack collective behavior. However, recent experimental observations have challenged this view. Phenomena such as strangeness enhancement~\cite{ALICE:2016fzo}, long-range ridge-like correlations~\cite{CMS:2011cqy}, and finite azimuthal anisotropies (flow coefficients) in high-multiplicity pp collisions have sparked intense discussion about the possible emergence of QGP-like effects even in small systems. More recently, heavy-ion-like effects have been found in low multiplicity pp collisions~\cite{ALICE:2023ulm,ALICE:2025bwp}, and probably at high jet-constituent multiplicities~\cite{CMS:2023iam,Vertesi:2024fwl}. These findings suggest that final-state collective dynamics which were previously thought to be only present in heavy-ion collisions, may also be present in high-multiplicity pp and p--Pb collisions.

A well-established approach to probe the UE and search for collective effects involves studying event activity as a function of leading-particle transverse momentum ($p_{\rm T}^{\text{trig}}$). Specifically, UE activity is quantified via charged-particle number density or summed $p_{\rm T}$ in the three topological regions. Experimental studies for pp collisions at RHIC~\cite{STAR:2019cie} and LHC~\cite{CMS:2010rux, CMS:2012oqb, ATLAS:2010kmf,ALICE:2011ac} have demonstrated that UE activity shows a sharp rise with increasing $p_{\rm T}^{\text{trig}}$, followed by saturation at high values in the transverse region~\cite{ALICE:2019mmy}. This feature can be interpreted as the transition to a regime dominated by MPI~\cite{CDF:2001onq}. Furthermore, UE activity in the saturation region has been found to grow more rapidly with increasing center-of-mass energy~\cite{ATLAS:2017blj,ALICE:2022fnb}, and the multiplicity distributions exhibit KNO-like scaling properties~\cite{Ortiz:2017jaz}. However, it is important to study  the jet-like signals and disentangle the effect of medium-like contribution, which can give us information about possible medium-like effects in small systems.

While UE properties in pp collisions are well characterized~\cite{STAR:2019cie}, relatively few studies exist for proton--nucleus (p--Pb) systems. This limits our understanding of the role of nuclear effects in UE dynamics. A recent study by ALICE~\cite{ALICE:2022fnb} found that the particle density in the transverse region saturates at $p_{\rm T}^{\rm trig} \approx 5$\,GeV/$c$ suggesting a bias on the nucleon-nucleon impact parameter, and to some extend, on the p--Pb impact parameter. Moreover, a notable dip in the ratio of jet-like yields between pp and p--Pb collisions at low $p_{\rm T}^{\text{trig}}$ was observed, suggesting the possible influence of medium-like effects or collective behavior even in small asymmetric systems. Understanding such features requires dedicated modeling and comparison across different theoretical frameworks. 

In order to better understand the UE activity and jet-medium interaction,  three state-of-the-art event generators are used to simulate pp collisions: \textsc{Pythia8}, the AMPT (A Multi-Phase Transport) model, and EPOS4, while for p-Pb collisions, AMPT is employed to simulate the collisions. Each of these generators implements different physics mechanisms relevant to UE modeling and effects that mimic thermalization:

\begin{itemize}
    \item \textsc{Pythia8} is widely used for simulating pp collisions and incorporates a comprehensive treatment of QCD processes, including MPI and color reconnection (CR)~\cite{Sjostrand:2014zea}. CR in particular is capable of mimicking collectivity-like behavior through string length minimization, which can explain features such as strangeness enhancement and ridge-like correlations in high-multiplicity pp events~\cite{OrtizVelasquez:2013ofg}.
    
    \item AMPT is a hybrid model particularly successful in describing collective flow in heavy-ion collisions~\cite{Lin:2004en,Zhang:2019utb}. It includes partonic scatterings and hadronization via quark coalescence in the string melting (SM) version. While it has been extensively used in nucleus--nucleus studies, its application to pp system remains less explored. The inclusion of the string melting mechanism makes it a suitable tool to investigate equilibrium-like effects in small systems.

    \item EPOS4 is a recently updated version of the EPOS model that includes a viscous (3+1)D hydrodynamic evolution for dense regions of the collision. It adopts a core-corona hadronization scheme and is capable of reproducing features associated with QGP formation~\cite{Werner:2023zvo}. EPOS4 provides a valuable reference to study hydrodynamic behavior and final-state interactions even in small collision systems.
\end{itemize}

In this article, the underlying-event activity is studied in the three topological regions using these three event generators for pp collisions, whereas we use only AMPT to simulate the p--Pb collisions. Their predictions are compared with experimental data to assess their ability to reproduce observed features and to explore the role of partonic scatterings, hydrodynamics, and color reconnection in shaping UE dynamics. The paper is organized as follows: in Section~\ref{sec2}, event generators and simulation configurations are presented. Section~\ref{sec3} outlines the methodology and observables used in this analysis. In Section~\ref{results}, the results are presented and discussed. Finally, Section~\ref{summary} summarizes our findings and outlines potential implications for future studies.

\section{Event generation}
\label{sec2}
\subsection{PYTHIA8}

To simulate ultrarelativistic collisions such as electron-electron, electron-positron, and proton-proton collisions, event generators like \textsc{Pythia8} are important tools in high-energy physics. \textsc{Pythia8} has demonstrated remarkable success in reproducing LHC experimental data~\cite{Skands:2014pea,Bierlich:2015rha,CMS:2010ifv}. It provides a comprehensive simulation framework for both QCD and QED processes, including hard scatterings (described using matrix elements), soft interactions, parton distribution functions (PDFs), and both initial- and final-state parton showers. A key feature of \textsc{Pythia8} is its treatment of multiparton interactions, which are essential for modeling the UE, which are a significant source of soft particle production accompanying the hard scattering.

The hadronization process is handled via the Lund string fragmentation model, which converts colored partons into color-neutral hadrons. Another critical ingredient is color reconnection, whereby partons from different MPI systems can form new color strings based on proximity in momentum space. This mechanism is particularly important in explaining collective-like effects observed in high-multiplicity small systems, such as enhanced particle production at intermediate $p_{\rm T}$ and the emergence of long-range correlations. \textsc{Pythia8} also simulates resonance decays, beam remnants, and limited final-state rescattering. The color reconnection parameters are tunable and play a significant role in describing phenomena like strangeness enhancement~\cite{Nayak:2018xip} and the ridge structure observed in pp collisions. These features make \textsc{Pythia8} a versatile and powerful tool for investigating both perturbative and nonperturbative aspects of QCD in small systems.

\subsection{A Multi-Phase Transport Model}

The AMPT model~\cite{Lin:2004en} is a hybrid framework developed to simulate nuclear and hadronic collisions across a broad energy range. It is particularly well-suited for studying the dynamical evolution of matter created in heavy-ion and small-system collisions. AMPT consists of four main components. First, the initial conditions are generated using the HIJING model, which calculates the minijet production cross-sections in nucleon-nucleon collisions and embeds them within a Glauber geometry for larger systems. Second, partonic scatterings are described by Zhang’s Parton Cascade (ZPC)~\cite{Zhang:1997ej}, which solves the Boltzmann equation for elastic parton-parton collisions.

In this study, we employ the string melting version of AMPT, in which all excited strings are converted into partons before hadronization. The third stage involves hadronization via a spatial quark coalescence mechanism, which combines nearby partons in coordinate space into hadrons based on their quantum numbers. This process has been shown to better reproduce collective flow observables and intermediate-$p_{\rm T}$ spectra compared to string fragmentation. Finally, the produced hadrons undergo further interactions in a hadronic transport phase, which includes baryon-baryon, meson-baryon, and meson-meson scatterings modeled using a relativistic transport framework.

AMPT is capable of simulating pp, pA, and AA collisions and is particularly effective at describing the transition from partonic to hadronic matter. In this work, the version 2.26t9b with the default parameters is used. The model's multi-phase structure enables detailed studies of the time evolution of the system, making it a valuable tool for exploring both the collective behavior and microscopic interactions in relativistic collisions.

\subsection{EPOS4}

EPOS4 is the latest version of the EPOS (Energy-conserving quantum mechanical multiple scattering approach based on Partons, Off-shell remnants, and Splitting of parton ladders) Monte Carlo event generator~\cite{Werner:2023jps}. It introduces a sophisticated parallel multiple-scattering formalism rooted in a unified theoretical approach that blends Gribov-Regge theory with perturbative QCD~\cite{Drescher:2000ha}, including DGLAP evolution and saturation effects. This framework allows EPOS4 to consistently describe soft and hard processes, as well as the interplay between them, within the same event.

EPOS4 incorporates energy-momentum conservation at every scattering vertex and implements AGK (Abramovsky–Gribov–Kancheli) cutting rules for multiple scattering. The partonic stage is followed by a core-corona separation at an early proper time $\tau_0 = 0.4$~fm/$c$. The “core” region, characterized by a high density of overlapping strings, undergoes collective evolution using (3+1)D viscous hydrodynamics through the vHLLE framework. The equation of state is taken from lattice QCD calculations, assuming a smooth crossover transition from the deconfined to the hadronic phase. Statistical hadronization occurs when the energy density drops below the critical energy density for hadronization, $\varepsilon_{\rm H} \simeq 0.57~\rm GeV/fm^3$.

The “corona” region, in contrast, hadronizes via Lund string fragmentation and consists of high-$p_{\rm T}$ particles that escape the dense core region. Notably, EPOS4 allows for interactions between core and corona hadrons in the late stages, simulating rescattering effects through the UrQMD hadronic cascade~\cite{Bleicher:1999xi}. This hybrid treatment enables EPOS4 to reproduce a wide range of observables in both small and large systems, including flow-like signatures in high-multiplicity pp and pA events, thereby making it a robust and theoretically grounded tool for studying collective effects across different collision systems.

\section{Methodology}
\label{sec3}

The event activity in each topological region is quantified by the primary charged particle number density which can be calculated as a function of $p_{\rm T}^{\rm trig}$:

\begin{equation}
    \bigg\langle \frac{{\rm d}^{2}N_{\rm ch}}{\rm{d}\eta \rm{d}\varphi}\bigg\rangle (p_{\rm T}^{\rm trig}) = \frac{1}{\Delta \eta \Delta \varphi} \frac{1}{N_{\rm ev}(p_{\rm T}^{\rm trig})}N_{\rm ch}(p_{\rm T}^{\rm trig})
\end{equation}
where, $N_{\rm ev}(p_{\rm T}^{\rm trig})$ is the total number of events with the leading particle in a given $p_{\rm T}^{\rm trig}$ interval. One can also look for the charged-particle number density in the jet-like signal, which is calculated from the diﬀerence between the number density in the toward (or away) region to that of the transverse region.

\begin{multline}
    \bigg \langle \frac{{\rm d}^{2}N_{\rm ch}}{\rm{d}\eta \, \rm{d}\varphi} \bigg \rangle^{\rm jet~toward(away)}(p_{\rm T}^{\rm trig}) \\
    = \bigg [  
        \bigg \langle \frac{{\rm d}^{2}N_{\rm ch}}{\rm{d}\eta \, \rm{d}\varphi} \bigg \rangle^{\rm toward(away)} 
        - \bigg \langle \frac{{\rm d}^{2}N_{\rm ch}}{{\rm d}\eta \, {\rm d}\varphi} \bigg \rangle^{\rm transverse}
    \bigg ](p_{\rm T}^{\rm trig})
\end{multline}

This approach was followed by the ALICE experiment~\cite{ALICE:2022fnb}.

\section{Results and discussion}
\label{results}

\begin{figure*}[ht!]
    \centering
    \includegraphics[width = 1\linewidth]{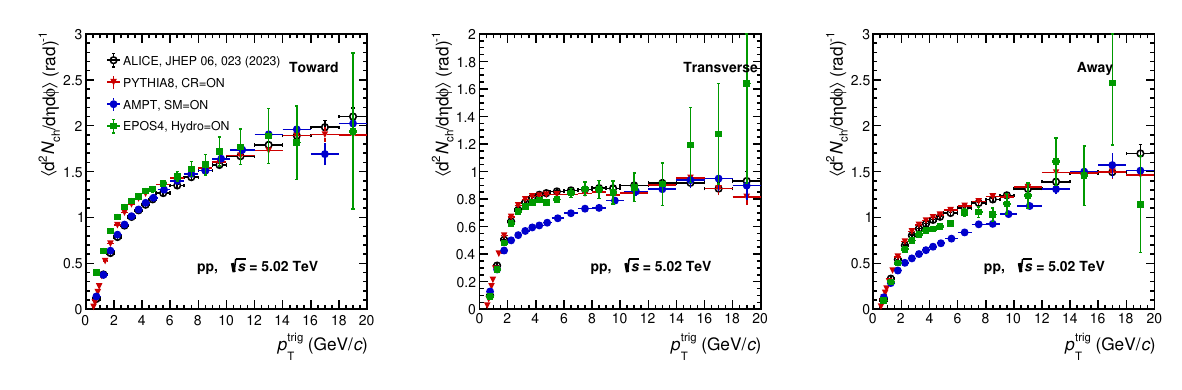}
    \caption{Charged-particle number density as a function of $p_{\rm T}^{\rm trig}$ measured in pp collisions at $\sqrt{s}$ = 5.02 TeV for three topological regions; toward side (left panel), transverse side (middle panel) and away side (right panel). We have compared results from PYTHIA8, AMPT and EPOS4 with the ALICE experimental data~\cite{ALICE:2022fnb}.}
        \label{fig2}
\end{figure*}

For the comparison with existing data, pp and p--Pb collisions at $\sqrt{s_{\rm NN}}$ = 5.02 TeV were simulated. The \textsc{Pythia8} sample consists of 10 million minimum-bias pp events with simulations performed for both color reconnection ON and OFF configurations. For the EPOS4 model, 1 million pp events have been simulated with both hydrodynamics ON and OFF. In the case of AMPT, 7 million pp events are simulated for string melting ON and OFF configuration as well as 3 million p--Pb events for both SM ON and OFF setups. 

Figure~\ref{fig2} displays the charged-particle number density as a function of \( p_{\rm T}^{\rm trig} \) across all three topological regions for pp collisions at center of mass energy 5.02 TeV. Results are presented for simulations that include medium-like effects (CR, hydrodynamics or string melting). A minimum transverse momentum threshold of 0.5~GeV/$c$ is applied to the associated particles, with a pseudorapidity cut of $|\eta| < 0.8$ on all particles. Results from \textsc{Pythia8}, EPOS4, and AMPT are shown and compared with data from the ALICE experiment. In the toward region, the results from the event generators generally agree well with ALICE data. In the transverse region (middle panel), a saturation of the charged-particle density is observed around $ p_{\rm T}^{\rm trig} \sim 5 $~GeV/$c$ for most models, indicating that the underlying event becomes largely independent of the leading particle's transverse momentum beyond this point. This saturation behavior is absent in AMPT. For the transverse and away regions, AMPT deviates significantly from experimental observations, underestimating the yields. This discrepancy arises from the event generation in AMPT, which relies on \textsc{HIJING} version 1.383. This version employs an older implementation of \textsc{Pythia} (v5.7), which lacks a full treatment of multi-parton interactions. In AMPT, the minijet partons produced from hard scatterings in HIJING model can lose energy via gulon splitting and trasfer it to nearby soft strings, which enhances the particle production not only in the toward and away regions but also in the transverse regions through re-energized strings~\cite{Lin:2004en}. Additionally, a higher $p_{\rm T}^{\rm trig}$ in AMPT can bias the system toward a denser partonic medium, leading to increased coalescence and large particle yields across all azimuthal regions, including the transverse—unlike in \textsc{Pythia8}, where jets are typically not quenched and underlying event activity saturates. On the other hand, both \textsc{Pythia8} and EPOS4 are largely consistent with ALICE data in all three regions. 

The color reconnection mechanism in \textsc{Pythia8} plays a crucial role in shaping UE observables, especially in the transverse region, which is most sensitive to soft QCD processes and multiple parton interactions. Enabling CR in PYTHIA results in a reduced charged-particle multiplicity ($N_{\rm ch}$) due to the formation of shorter, reconnected color strings, which fragment into fewer but generally harder particles~\cite{Ortiz:2018vgc}. In AMPT, the inclusion of the string melting mechanism significantly modifies the underlying event in pp collisions. When SM is enabled, excited strings are converted into a deconfined partonic medium that undergoes parton cascade interactions before hadronization. This leads to enhanced radial flow, broader azimuthal correlations, and potentially ridge-like structures in high-multiplicity events---phenomena typically associated with collective behavior in heavy-ion collisions~\cite{Ji:2023eqn,MenonKavumpadikkalRadhakrishnan:2023cik}. The resulting $p_{\rm T}$ spectra are harder, and the overall multiplicity can be reduced due to recombination of partons into fewer hadrons and conversion of energy into transverse expansion rather than particle production. This behavior is analogous to the effects observed with CR in \textsc{Pythia8}. Conversely, in the simulations without string melting, hadronization occurs directly via Lund string fragmentation without partonic rescattering. This leads to the production of more soft hadrons and higher multiplicities, resembling more traditional string fragmentation behavior with minimal collectivity. The EPOS4 model incorporates a (3+1)D viscous hydrodynamic phase when hydrodynamics is enabled. This significantly alters particle production and momentum distributions in all topological regions by introducing collective expansion and medium-modified hadronization. As a result, the transverse region activity in EPOS with hydrodynamics tends to be elevated, capturing features commonly associated with QGP-like behavior even in small systems such as pp collisions.

\begin{figure*}[ht!]

    \includegraphics[width = 0.32\linewidth]{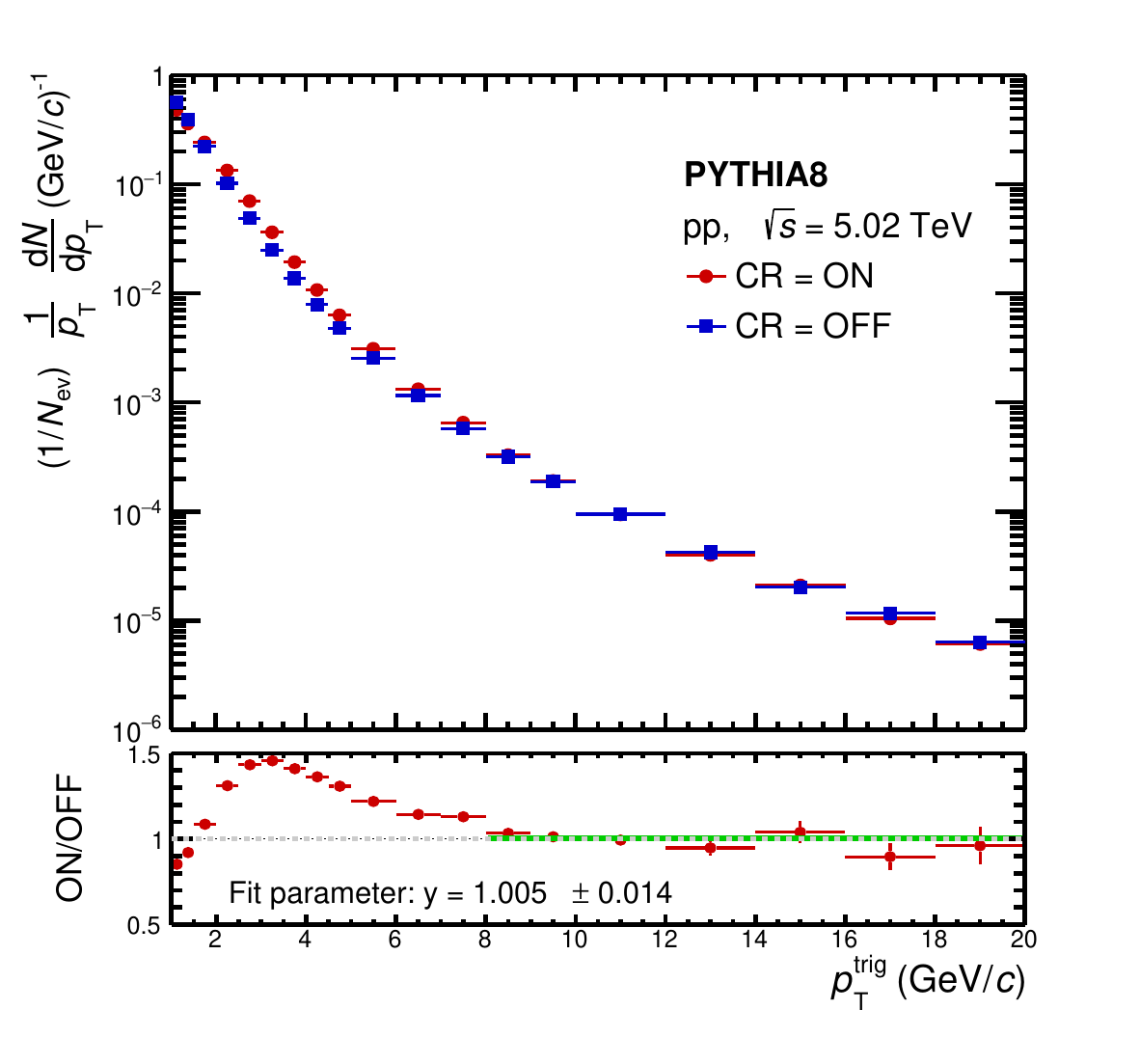}
        \includegraphics[width = 0.32\linewidth]{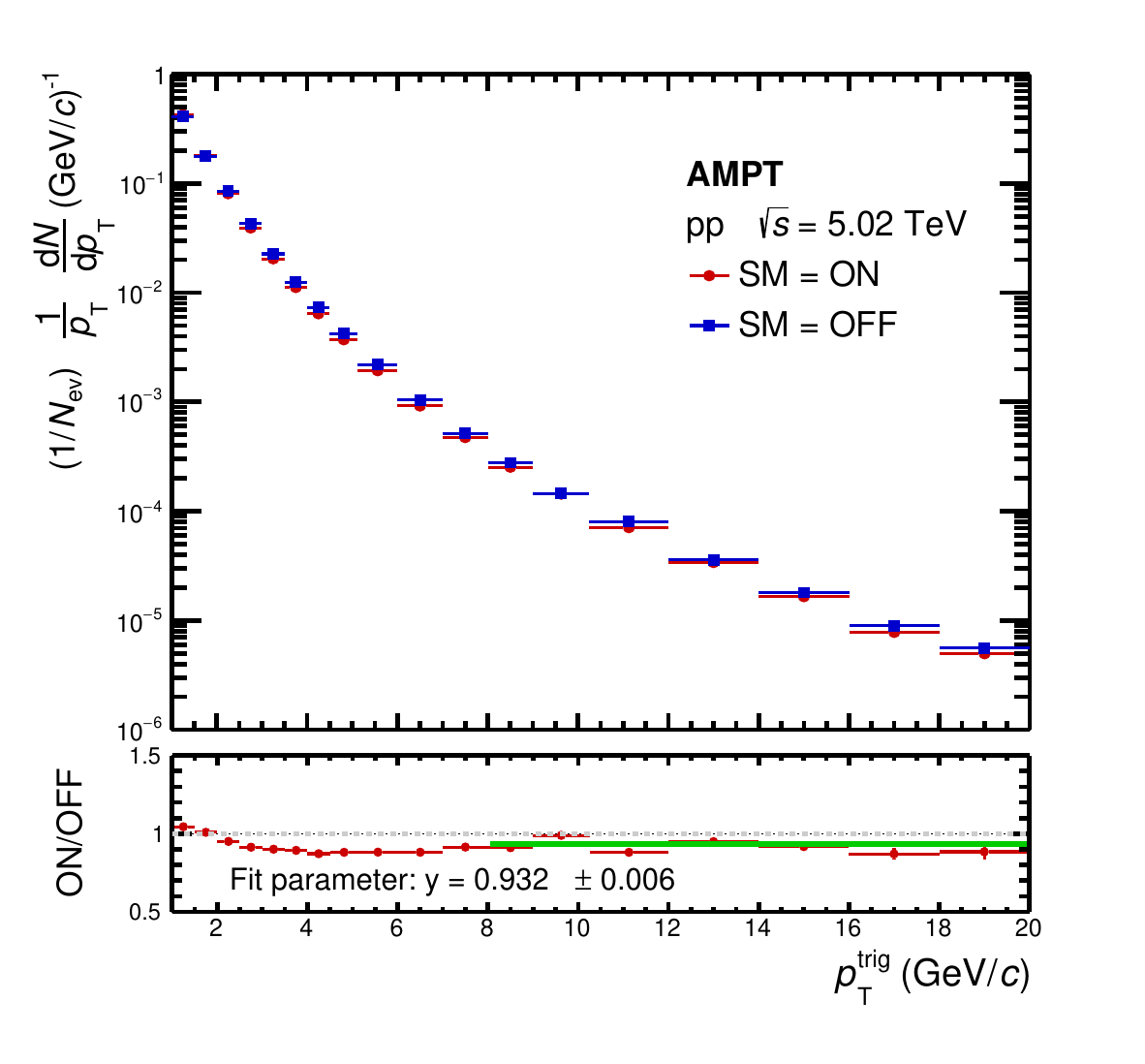}
            \includegraphics[width = 0.32\linewidth]{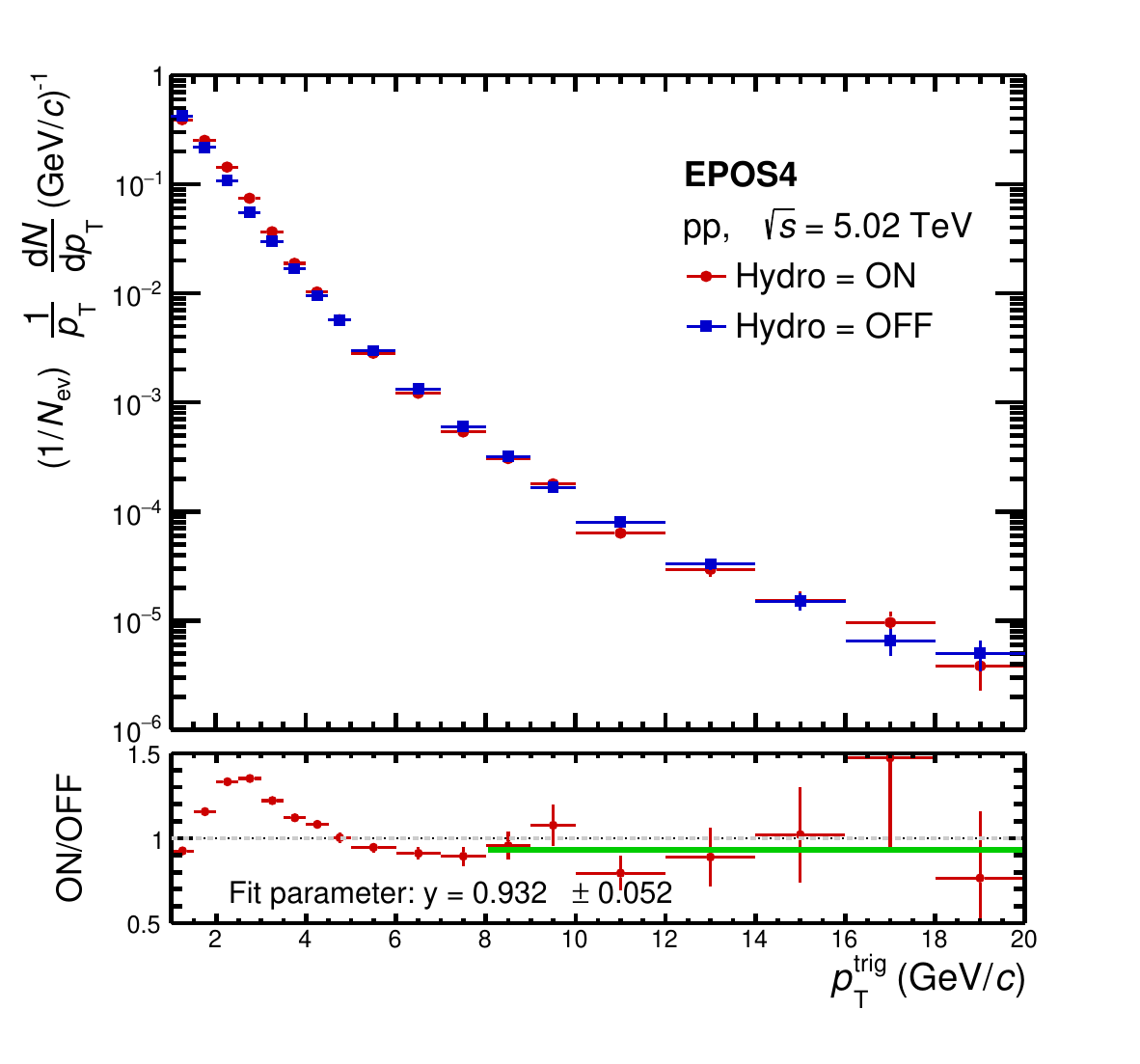}
    \caption{Leading particle transverse momentum spectra in pp collisions at $\sqrt{s}$ = 5.02 TeV for \textsc{Pythia8} (left panel), AMPT (middle panel) and EPOS4 (right panel). The green solid line in the lower pads show the fit to the ratio in the range $8 \leq p_{\rm T}^{\rm trig} \leq 20$ GeV/c.}
    \label{fig4}
\end{figure*}

\begin{figure*}[ht!]
        \includegraphics[width = 0.32\linewidth]{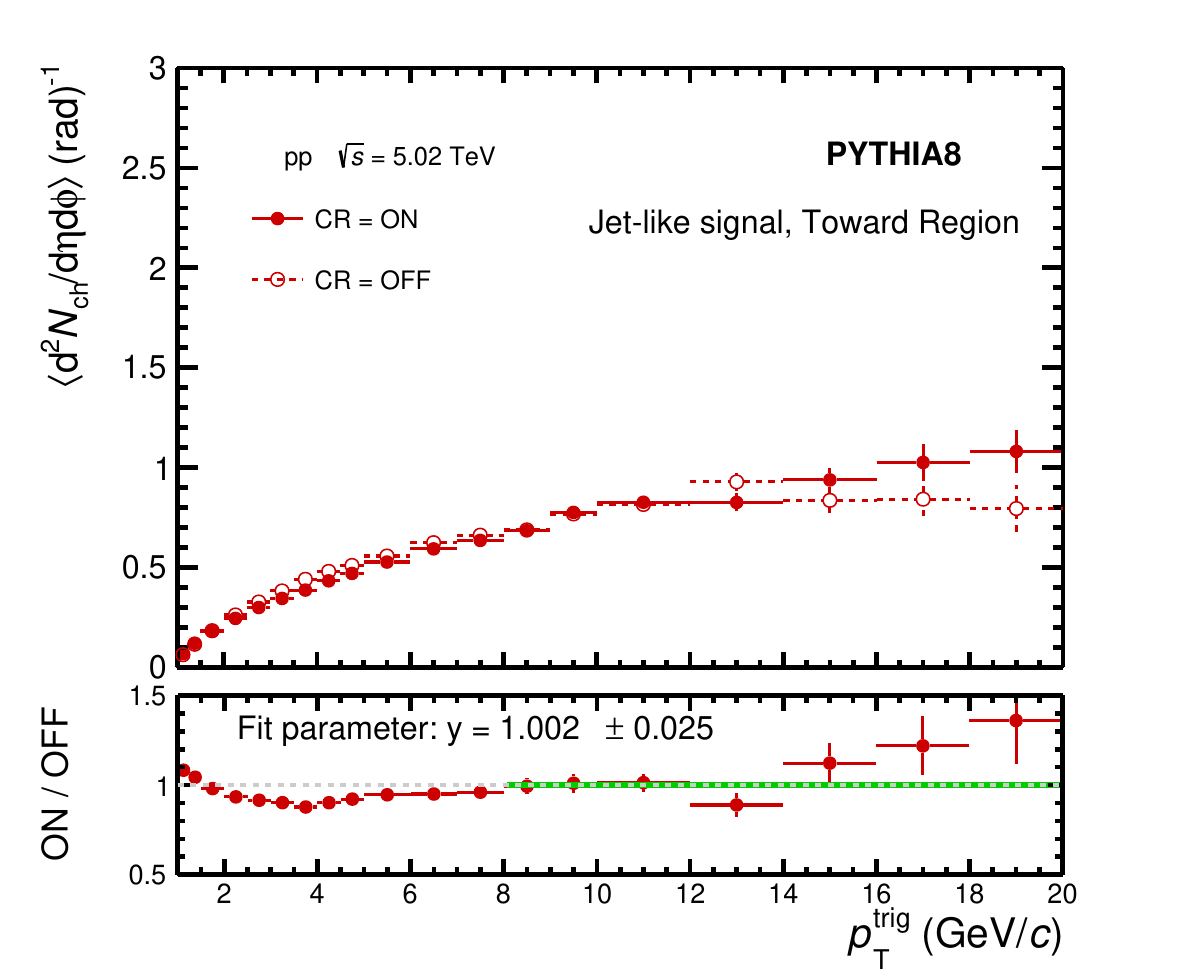}
    \includegraphics[width = 0.32\linewidth]{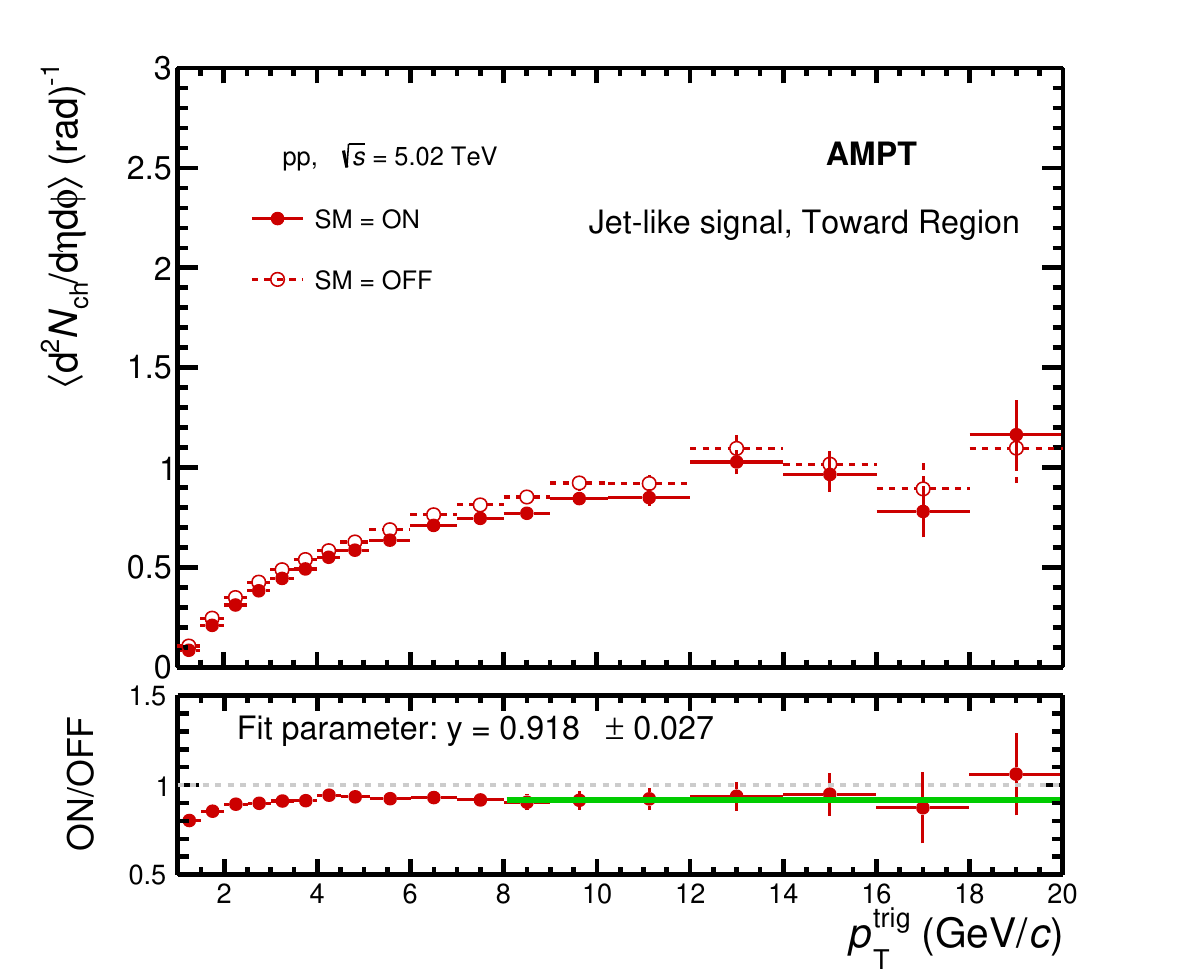}
        \includegraphics[width = 0.32\linewidth]{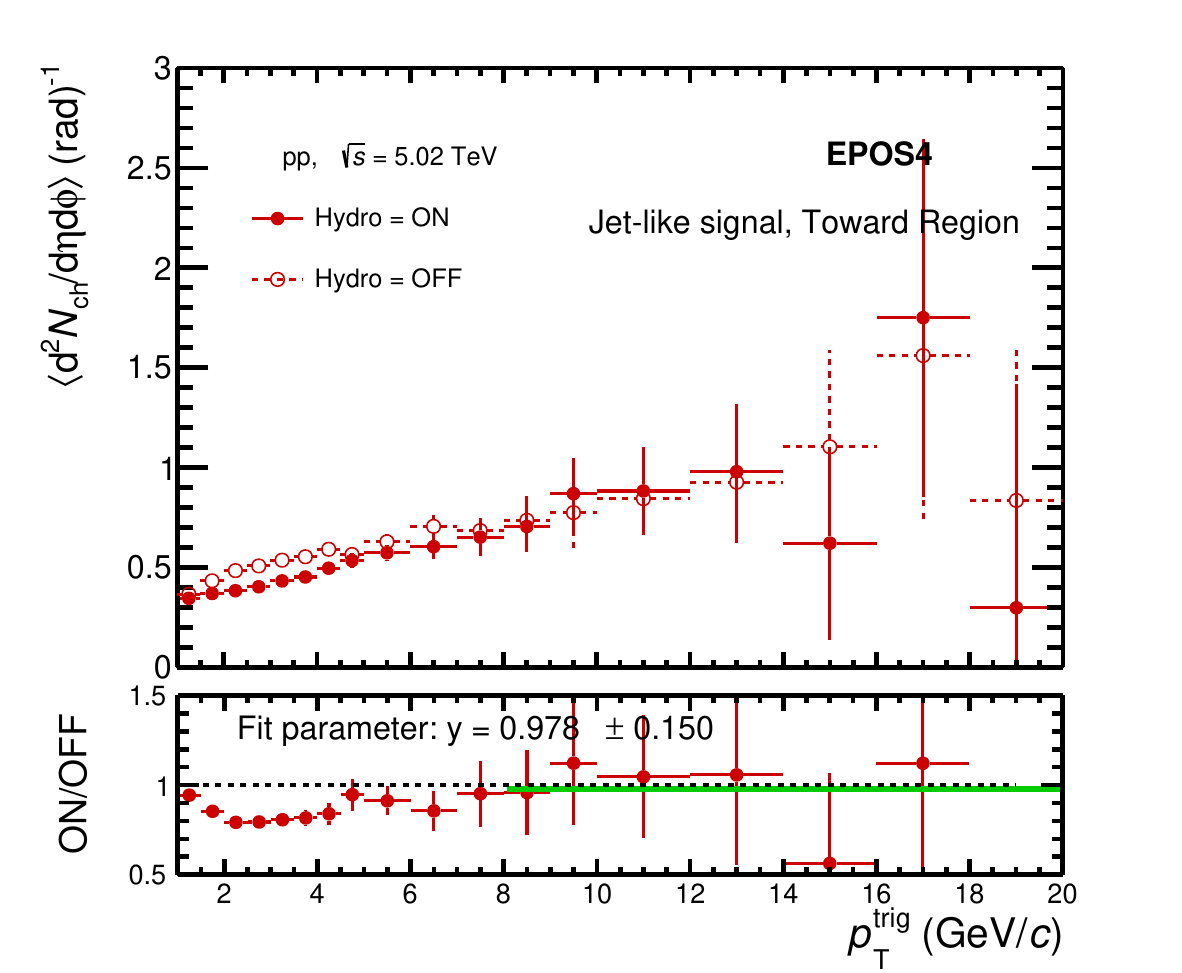}
    
    \caption{Charged-particle number density for jet-like signal in the toward region as a function of $p_{\rm T}^{\rm trig}$ for pp collisions at $\sqrt{s}$ = 5.02 TeV estimated from \textsc{Pythia8} (left panel), AMPT (middle panel) and EPOS4 (right panel). The green solid line in the lower pads show the fit to the ratio in the range $8 \leq p_{\rm T}^{\rm trig} \leq 20$ GeV/c.}
    \label{fig5}
\end{figure*}

The leading-particle transverse momentum spectra for pp collisions at center of mass energy 5.02 TeV using all three event generators is shown in Fig.~\ref{fig4}. The $p_{\rm T}^{\text{\rm trig}}$ spectra, representing the transverse momentum of the leading particle in each event, exhibit different behaviors across the three models. The left panel shows the event averaged particle distribution as a function of $p_{\rm T}^{\rm trig}$ using \textsc{Pythia8} with CR ON and OFF. The lower pad shows the ratio between CR = ON and CR = OFF, which we shall refer to as medium-like to vaccum ratio. In \textsc{Pythia8}, enabling color reconnection leads to a harder $p_{\rm T}^{\rm trig}$ distribution in the intermediate region (2--8 GeV/$c$), resulting in a ratio greater than one. This is due to CR introducing collective-like effects that boost the momenta of soft partons, thereby increasing the likelihood of producing a harder leading particle. At higher $p_{\rm T}^{\rm trig}$, the effect of CR diminishes, and the ratio becomes unity. To verify this, a zeroth order polynomial is fitted to the ratio in the range $p_{\rm T}^{\rm trig}$ = 8--20 GeV/c. The fit shows that the ratio is unity within uncertainty. A similar trend is observed in EPOS4, shown in the right panel of Fig.~\ref{fig4}, where the inclusion of hydrodynamic evolution enhances soft particle production via radial flow, again leading to a harder $p_{\rm T}^{\rm trig}$ distribution and a ratio greater than one in the intermediate $p_{\rm T}$ region (2--5 GeV/$c$). At high $p_{\rm T}^{\rm trig}$, where particle production is dominated by hard scatterings, the hydro effect becomes negligible and the ratio flattens to one. In this case also, the fit shows that the ratio is consistent with unity within uncertainty with 1.31$\sigma$ deviation. 

In contrast, the AMPT model with string melting shows a consistently suppressed $p_{\rm T}^{\rm trig}$ spectrum, with the ratio (melting on / melting off) remaining slightly below one across the entire $p_{\rm T}$ range, as shown in the middle panel of Fig.~\ref{fig4}. In this case, the fit shows 11.33$\sigma$ deviation from unity. This behavior arises because the string melting mode, which involves partonic interactions and hadronization through coalescence, produces softer and more isotropic events, reducing the probability of having a single high-$p_{\rm T}$ leading particle. Even at higher $p_{\rm T}$, the lack of conventional jet fragmentation~\cite{Lin:2004en}  in this mode can lead to a softened leading particle spectrum as compared to the default setting.

Figure~\ref{fig5} shows the charged-particle multiplicity associated with jet-like signals in the toward region for pp collisions at $\sqrt{s} = 5.02$ ~TeV. The left panel shows the results from \textsc{Pythia8} simulations for both CR=ON and CR=OFF scenarios. The middle panel presents the AMPT estimations for SM=ON and SM=OFF configurations, while the right panel corresponds to EPOS4 results with hydro ON and OFF. All three generators exhibit an increasing trend in charged-particle multiplicity with $p_{\rm T}^{\rm trig}$, reflecting the rising contribution of hard processes and collimated jet activity. The bottom panels of each plot show the medium-like to vaccum ratios. In the case of \textsc{Pythia8} and EPOS4, a modest dip is observed at low $p_{\rm T}^{\rm trig}$, followed by a gradual rise and saturation around \( p_{\rm T}^{\rm trig} \simeq 5\)--6~GeV/$c$, where the ratios approach unity. In the low $p_{\rm T}^{\rm trig}$ region, genuine mini-jets start contributing. CR may reconnect jet partons to soft ones, redistributing energy and reducing jet coherence. This smears the jet axis, hence degrading the signal in the toward region. Similarly, in the case of EPOS4, the interplay between flow and hard jets distorts the signal. On the other hand, the saturation after $p_{\rm T}^{\rm trig} \geq 8$ GeV/$c$ indicates that the effect of color reconnection or hydrodynamic expansion becomes less significant at high $p_{\rm T}^{\rm trig}$, where hard processes dominate and UE modifications play a reduced role. This is again confirmed by fitting the ratio in the range $p_{\rm T}^{\rm trig}$ = 8--20 GeV/c. Both \textsc{Pythia8} and EPOS4 results show the ratio is consistent to unity within uncertainties.

However, AMPT shows a different trend; the ratio starts from below unity, slowly increasing trend with $p_{\rm T}^{\rm trig}$ but never approaching the unity, with the fit to the ratio showing a 3.04$\sigma$ deviation. This suppression arises because, in the string melting scenario, partons undergo significant rescatterings before hadronizing via quark coalescence, which tends to dilute the jets. In contrast, the default AMPT preserves jet-like features more effectively due to direct string fragmentation. The suppression indicates that even at high $p_{\rm T}$, the collective partonic interactions and the coalescence mechanism in the SM version continue to affect the jet-like signals. This is because, unlike in \textsc{Pythia8} or EPOS4, there is no clean separation between soft/hard processes or core/corona. Coalescence always modifies the hadron yield and correlation patterns, even at high $p_{\rm T}$. All three event generator studies, however, suggest us that the jet-like signals are significantly modified below $p_{\rm T}^{\rm trig} \leq 6$ GeV/$c$, and one should be cautious while doing trigger studies in that region. This ratio resembles the observations from ratio of pp to p--Pb collisions~\cite{ALICE:2022fnb}, where the jet-like yield in p--Pb normalized to that in pp collisions (``medium-like'' to ``vacuum-like'' ratio) exhibits a similar trend.

\begin{figure*}[ht!]
    \centering
    \includegraphics[width = 1\linewidth]{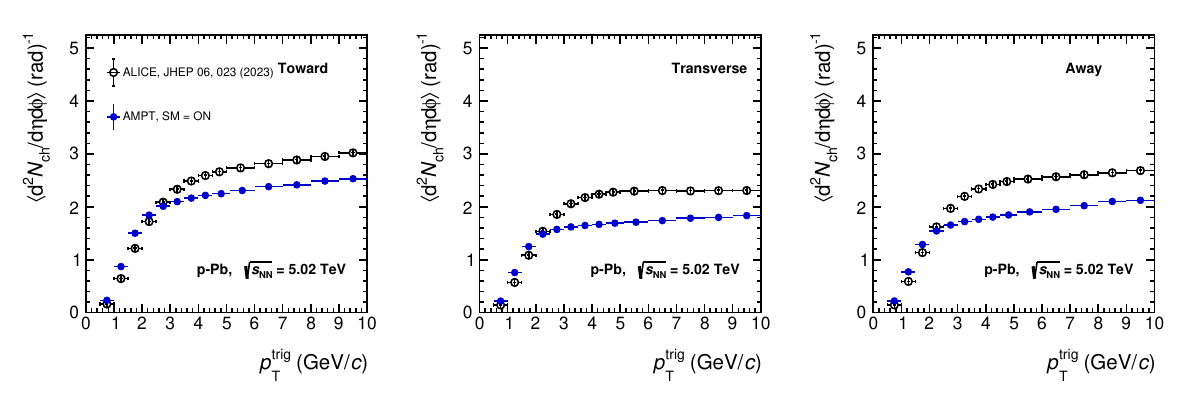}
    \caption{Charged-particle number density as a function of $p_{\rm T}^{\rm trig}$ measured in p--Pb collisions at $\sqrt{s_{\rm NN}}$ = 5.02 TeV for three topological regions; toward side (left panel), transverse side (middle panel) and away side (right panel). We have compared results from AMPT with the ALICE experimental data~\cite{ALICE:2022fnb}.}
    \label{fig3}
\end{figure*}

\begin{figure*}[ht!]
    \includegraphics[width = 0.49\linewidth]{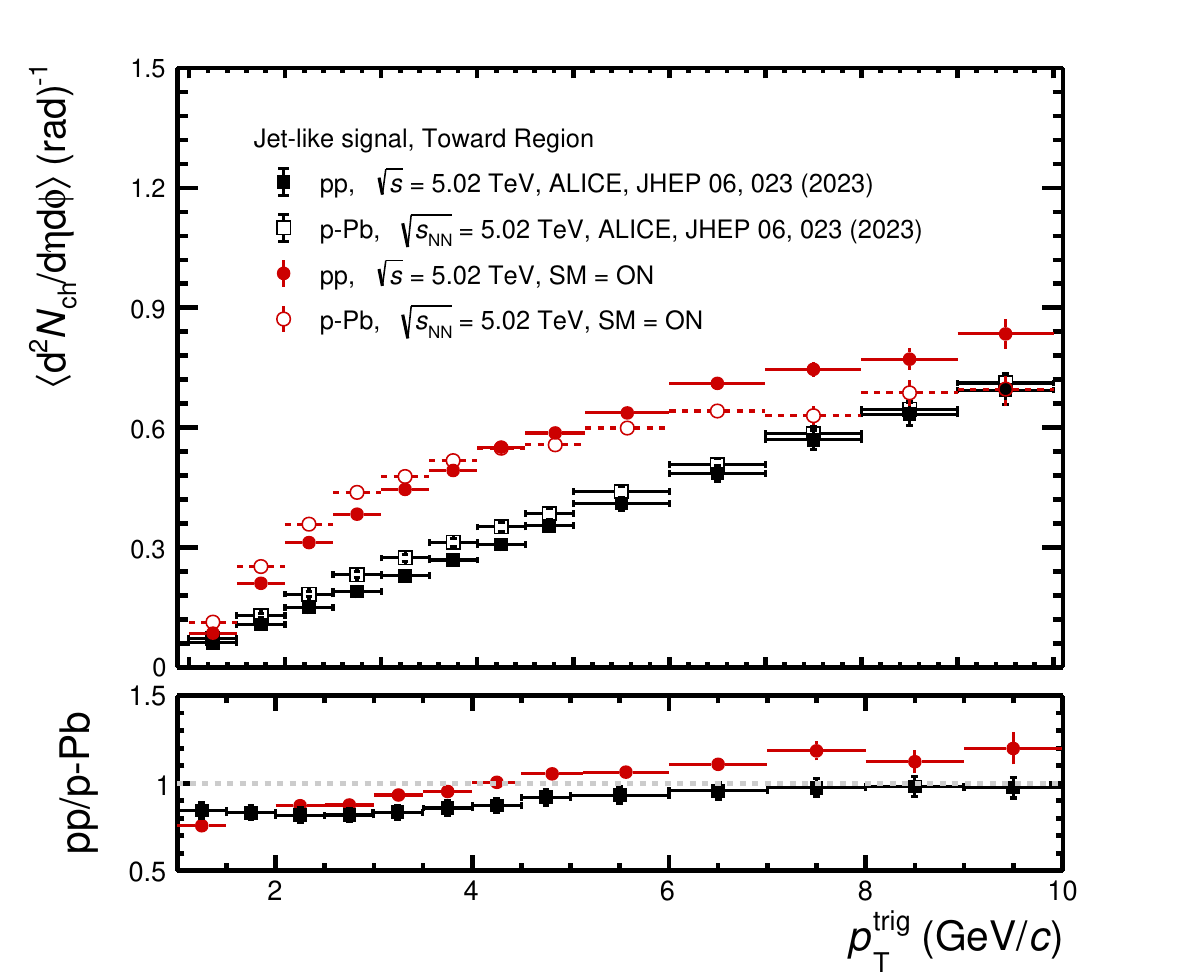}
        \includegraphics[width = 0.49\linewidth]{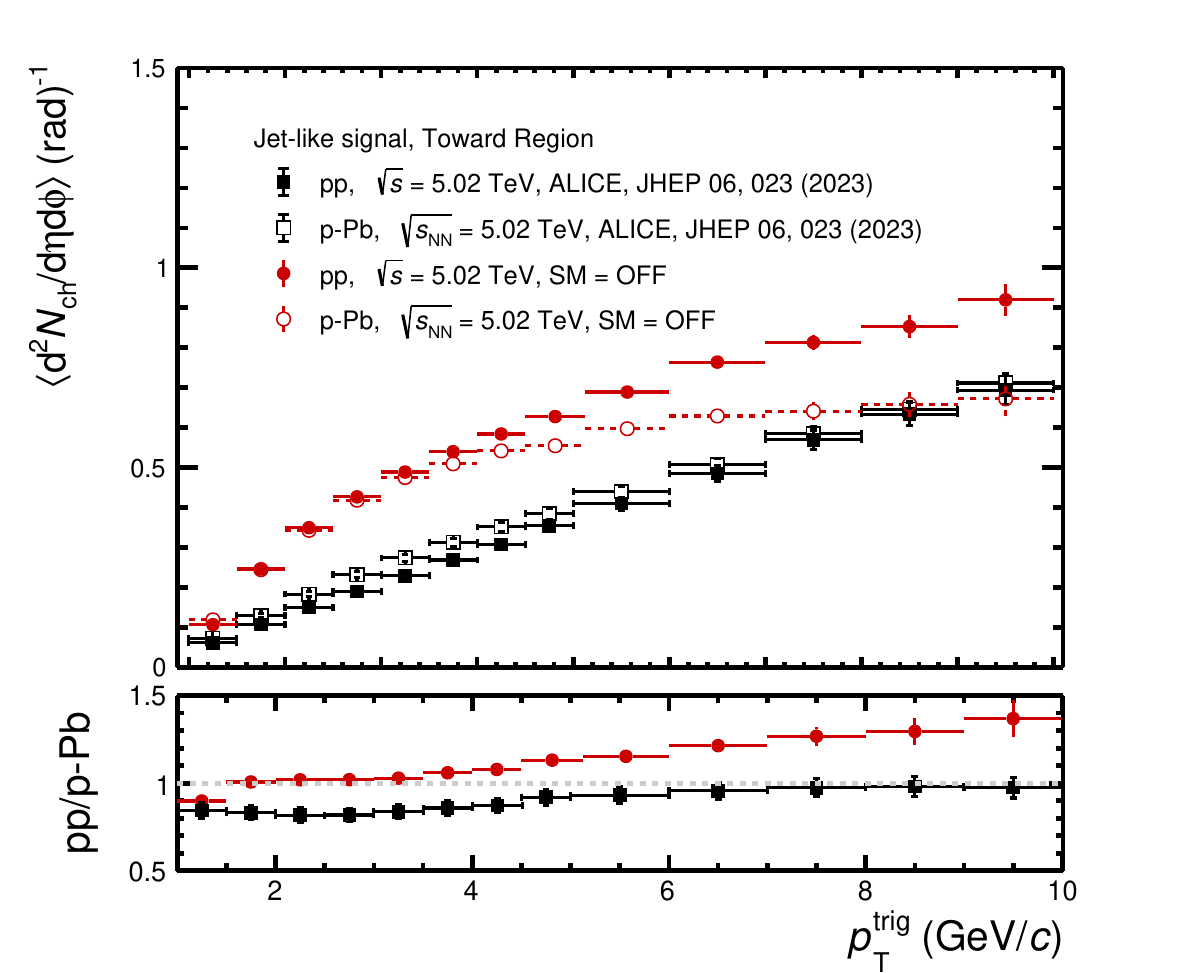}
    \caption{(Left panel) Charged-particle number desities for jet-like signal in the toward region as a function of $p_{\rm T}^{\rm trig}$ using AMPT string melting on for pp and p--Pb shown by red circles closed and open respectively. (Right panel) The same using AMPT string melting off. The black square markers are for ALICE experimental data, closed markers for pp and open markers for p--Pb~\cite{ALICE:2022fnb}.}
    \label{fig6}
\end{figure*}

The study is extended  to p--Pb collision, Fig.~\ref{fig3} shows the charged-particle multiplicity for p--Pb collisions at a center-of-mass energy of 5.02~TeV per nucleon pair, as estimated using the AMPT model across all three topological regions. These results are compared with experimental measurements from the ALICE collaboration. 
While the general trend of increasing multiplicity with trigger particle $p_{\rm T}^{\rm trig}$ is qualitatively reproduced by AMPT, the model significantly underestimates the absolute values of charged-particle multiplicity in all three regions. Similar underestimations have been reported in other models such as \textsc{Pythia} Angantyr and EPOS LHC for p--Pb collisions at the same energy~\cite{ALICE:2022fnb}. However, it is to be noted that the particle density in p--Pb collisions are higher than the pp collisions in their respective topological regions, which is trivially expected.

Finally, Fig.~\ref{fig6} shows the charged particle multiplicity of the jet-like signal in the toward region in both pp and p--Pb collisions for center of mass energies 5.02 TeV, where we have compared the estimations from AMPT with the ALICE experimental data. The left panel of Fig.~\ref{fig6} shows the results from the string melting ON configuration. A clear dependence on $p_{\rm T}^{\text{trig}}$ is observed in the pp to p--Pb ratio of jet-like yields. At low $p_{\rm T}^{\text{trig}}$ ( $\leq$ 5~GeV/$c$), the jet-like yield in p--Pb exceeds that in pp, resulting in a ratio less than unity. This trend matched very well with the observation from ALICE. This can be attributed to enhanced soft activity and mini-jet-like correlations in p--Pb due to the presence of multiple parton scatterings and possible collective behavior. The denser medium in p--Pb, even in the absence of full jet fragmentation, leads to an increased number of correlated particles in the toward region relative to the transverse region. In contrast, pp collisions, being more dilute, have weaker underlying event activity and fewer correlated soft particles at low $p_{\rm T}$, leading to smaller jet-like yields. However, this trend reverses at higher $p_{\rm T}^{\text{trig}}$ ($> 5$~GeV/$c$), where pp collisions begin to dominate in jet-like yield. The results in the high-$p_{\rm T}^{\rm trig}$ region should be taken with some caution, as there is disagreement between AMPT and data in pp collisions, which can contaminate such results. In any case, this transition of ratio reflects the onset of genuine hard-scattering processes. In the presence of a medium, as in p--Pb, partonic interactions in AMPT can lead to jet broadening, energy loss, and redistribution of fragments, thereby reducing the multiplicity contrast between the toward and transverse regions~\cite{Luo:2021hoo,Gao:2016ldo}. Meanwhile, in pp, the relatively clean environment allows high-$p_{\rm T}$ jets to remain collimated and distinct from the underlying event, enhancing the jet-like yield after background subtraction. This behavior is consistent with the expectation that partonic collectivity increases soft correlations in p--Pb at low $p_{\rm T}$, while jet degradation effects become more prominent at higher $p_{\rm T}$.

The right panel of Fig.~\ref{fig6} shows the charged particle multiplicity of the jet-like signal in the toward region, where we compare the AMPT string melting off version with ALICE experimental data. Here, the pp to p--Pb ratio of jet-like yield remains close to unity across the low $p_{\rm T}^{\text{trig}}$ range ($\leq$ 5~GeV/$c$), suggesting that in the absence of partonic collectivity, both systems exhibit similar soft particle production. Without parton-level scatterings, p--Pb does not show enhanced correlated particle production in the toward region beyond what is expected from the increased number of binary nucleon-nucleon collisions. At higher $p_{\rm T}^{\text{trig}}$, however, pp again exhibits higher jet-like yield than p--Pb, likely due to less hadronic rescattering and a more pristine jet structure. The broader hadronic environment in p--Pb introduces additional smearing of jet fragments and increases the background in the transverse region, thereby reducing the jet-like yield obtained by subtraction.

\section{Summary}
\label{summary}

In this article, the underlying-event properties and jet-medium interaction in pp and p--Pb collisions at $\sqrt{s_{\rm NN}} = 5.02$~TeV are studied. For the pp collision system study, three event generators; \textsc{Pythia8}, AMPT, and EPOS4 are used, whereas for p--Pb collisionw, only the AMPT model is used. Charged-particle number density was analyzed in the toward, transverse, and away regions as a function of the leading particle’s transverse momentum, with a threshold of 0.5~GeV/$c$ at mid-rapidity. All models reproduce the general trend of increasing UE activity with $p_{\rm T}^{\rm trig}$. Saturation in the transverse region is observed around 5~GeV/$c$ for most models, except AMPT. AMPT also underestimates the multiplicities in p--Pb collisions in all three regions, this behavior is in line with PYTHIA8 and EPOS-LHC estimation of p--Pb collisions reported by the ALICE Collaboration.

The leading-particle $p_{\rm T}^{\rm trig}$ spectra in pp collisions show a harder distribution in the intermediate $p_{\rm T}$ range due to collective effects like color reconnection in \textsc{Pythia8} and hydrodynamics in EPOS4. On the other hand, The AMPT string melting mode yields a consistently softer $p_{\rm T}^{\rm trig}$ spectrum due to partonic interactions and coalescence, suppressing high-$p_{\rm T}$ leading particle production. Effects from color reconnection in \textsc{Pythia8} and hydrodynamic expansion in EPOS4 were analyzed. Both lead to notable changes in soft particle production, with medium-like to vaccum yield ratios showing a dip at low \(p_{\rm T}^{\rm trig}\) and saturation around 5--6~GeV/$c$, marking the transition from soft to hard processes. These findings highlight the role of partonic and collective effects in shaping UE dynamics even in small systems. The study of jet-like yields shows that in AMPT (SM = ON), p--Pb exhibits higher yields at low \(p_{\rm T}^{\rm trig}\), indicating enhanced soft activity and collectivity, while pp dominates at high \(p_{\rm T}\), where jets are more collimated. In the SM OFF scenario, the pp to p--Pb ratio remains close to unity at low \(p_{\rm T}\), reflecting minimal collectivity.

The study also demonstrates that triggers with $p_{\rm{T}}^{\rm{trig}} <$  5--6~\rm{GeV}/$c$ are heavily influenced by UE dynamics, including multi-parton interactions and color reconnection, thus complicating the isolation of possible medium-like effects in pp collisions. Saturation of UE activity above $p_{\mathrm{T}}^{\mathrm{trig}} \sim 5-6~ {\rm GeV}/c$ suggests this as a practical threshold for jet-medium studies. Future analyses should prioritize high-$ p_{\rm{T}}^{\rm{trig}}$ $ > 8~\rm{GeV}/c$ triggers to suppress UE biases, particularly in small-system QGP searches. Extending these studies to identified triggers (e.g., heavy-flavor hadrons) and higher energies could further constrain the interplay between hard probes and collective dynamics.

\section*{Acknowledgment}
Authors acknowledge the technical support from Luciano D\'iaz Gonz\'alez and Jes\'us Eduardo Murrieta Le\'on. This work has been supported by DGAPA-UNAM PAPIIT No. IG100524 and PAPIME No. PE100124.

\bibliographystyle{utphys}
\bibliography{uepaper-refs}

\end{document}